\documentclass{PoS}
\usepackage{mathrsfs}

\usepackage{relsize}
\def\babar{\mbox{\slshape B\kern-0.1em{\smaller A}\kern-0.1em B\kern-0.1em{\smaller A\kern-0.2em R}}}

\title{Measurement of $B^+ \rightarrow K^+  \tau^+ \tau^-$, $B \rightarrow K^* l^+ l^-$ and $B \rightarrow K \pi^+ \pi^- \gamma$ decays at \babar\ }

\ShortTitle{Measurement of $B^+ \rightarrow K^+  \tau^+ \tau^-$, $B \rightarrow K^* l^+ l^-$ and $B \rightarrow K \pi^+ \pi^- \gamma$ decays at \babar\ }

\author{\speaker{Benjamin Oberhof}\thanks{on behalf of the \babar\ collaboration}\\
        LNF INFN, Frascati, Italy.\\
        E-mail: \email{benjamin.oberhof@lnf.infn.it}}

\abstract{
We present some recent measurements of rare flavor-changing neutral current $B$ decays, using data collected with the \babar\ detector at the PEP-II $e^+e^-$ collider at SLAC. 
First, we search for the rare process $B^+ \rightarrow K^+ \tau^+ \tau^-$ 
and we do not find evidence for a signal. 
The measured branching fraction is $(1.31^{+0.66} _{-0.61}$(stat.)$ ^{+0.35} _{-0.25}$(sys.)$) \times 10 ^{-3}$ with an upper limit, 
at the 90\% confidence level, of $\mathcal B (B \rightarrow K^+ \tau^+ \tau^-)< 2.25 \times 10^{-3}$. 
We then study the lepton forward-backward asymmetry $\mathcal A_{FB}$ and the longitudinal $K^*$ polarization $F_L$ in the rare decays $B \rightarrow K^* l^+ l^-$, where $l^+ l^-$ is either $e^+e^-$ or $\mu^+ \mu^-$. We 
report results for both the $K^{*}(892)^0 l^+ l^-$ and $K^{*}(892)^+ l^+ l^-$ final states, as well as their combination $K^* l^+ l^-$, in five disjoint dilepton mass-squared bins. 
Finally, we measure the time-dependent CP asymmetry in the radiative-penguin decay $B^0 \rightarrow K^0_S \pi^- \pi^+ \gamma$. 
The $K \pi \pi$ resonant structure is extracted by an amplitude analysis of the $m_{K\pi\pi}$ and $m_{K\pi}$ spectra in $B^+ \rightarrow K^+ \pi^- \pi^+ \gamma$ decays. 
We use these results to extract the mixing-induced CP parameters of the process $B^0 \rightarrow K_S ^0  \rho ^0 \gamma$ from the time-dependent
analysis of $B^0 \rightarrow K_S ^0 \pi^+ \pi^- \gamma$ decays 
and obtain $S_ {K^0_S \rho ^0 \gamma} = -0.18 \pm 0.32 $(stat.)$^{+0.06} _{-0.05}$(syst.). 
}

\FullConference{XIII International Conference on Heavy Quarks and Leptons\\
		22-27 May, 2016\\
		Blacksburg, Virginia, USA}

\begin{document}

\section{Introduction}
Flavor-changing neutral current (FCNC) B decays of the form $B \rightarrow K^{(*)}  X $ where $X= ll$ or $\pi \pi \gamma$ and $l = e, \mu$ are highly suppressed in the Standard Model (SM). 
The lowest-order SM processes contributing to these decays are the photon and $Z$ penguins and the $W^+ W^-$ box diagrams. These decays can provide a stringent test of the SM and a fertile ground for New Physics (NP) searches as virtual particles may enter in the loop and allow us to probe new physics at large mass scales. Details and references for each of the decay modes covered in this work are given in the following sections. 

We use data recorded by the \babar\ detector at the PEP-II asymmetric-energy $e^+e^-$ storage rings operated at the SLAC National Accelerator Laboratory. 
The data sample consists of 424 fb$^{-1}$  of $e ^+ e^-$ collisions recorded at the center-of-mass (CM) energy $\sqrt s = 10.58$ GeV. 
The cross section for $B \bar B$-pair production at the $\Upsilon(4S)$ is $\sigma _{B \bar B} \sim 1.1$ nb corresponding to a data sample of about $471 \times 10^6$ $B \bar B$-pairs \cite{lumi}. 
A detailed description of the \babar\ detector is given elsewhere \cite{detector}.

\section{Search for $B^+ \rightarrow K^+  \tau^+ \tau^-$}
\begin{figure*}[!htb]
\centering
\includegraphics[width=0.48\textwidth]{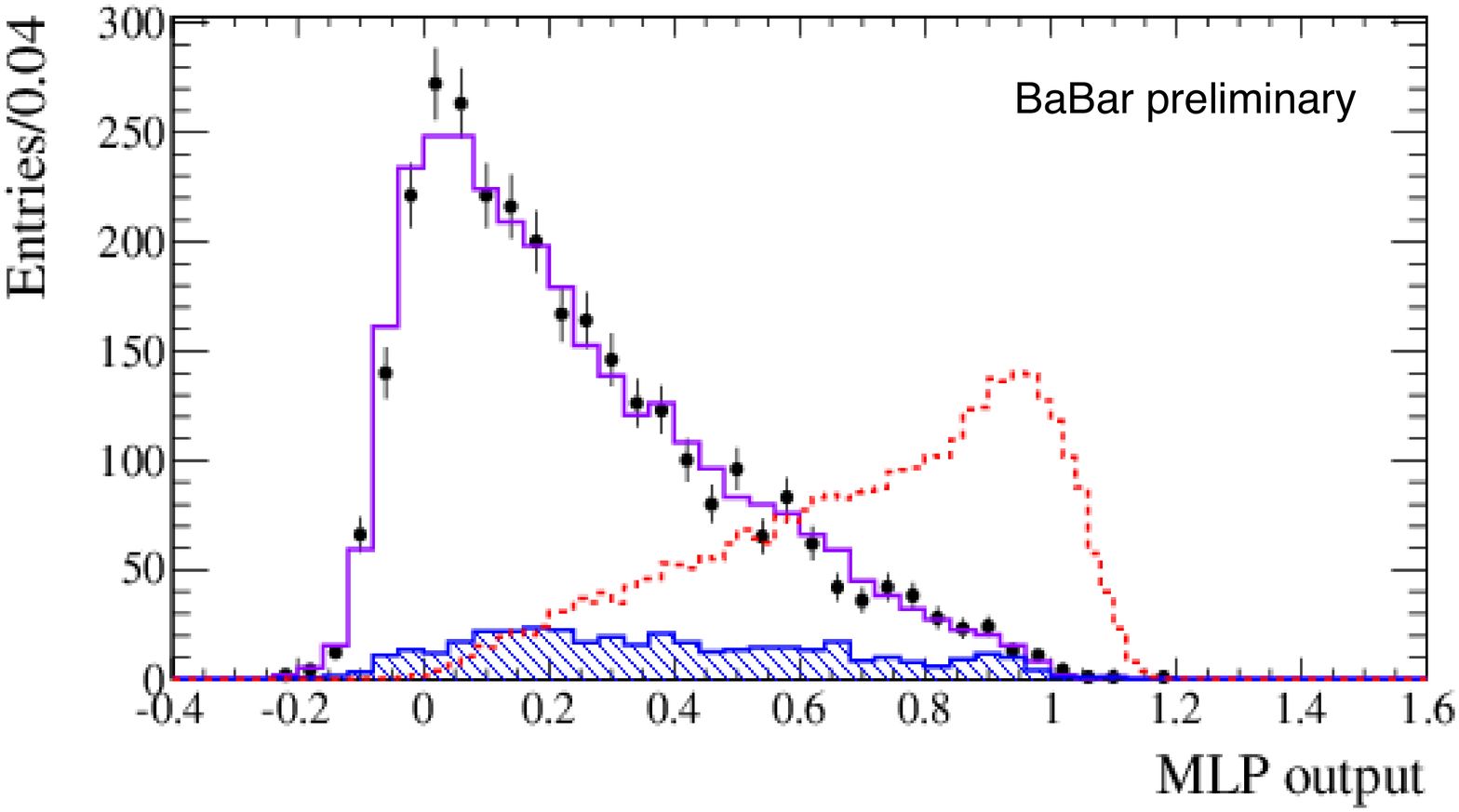}
\quad
\includegraphics[width=0.48\textwidth]{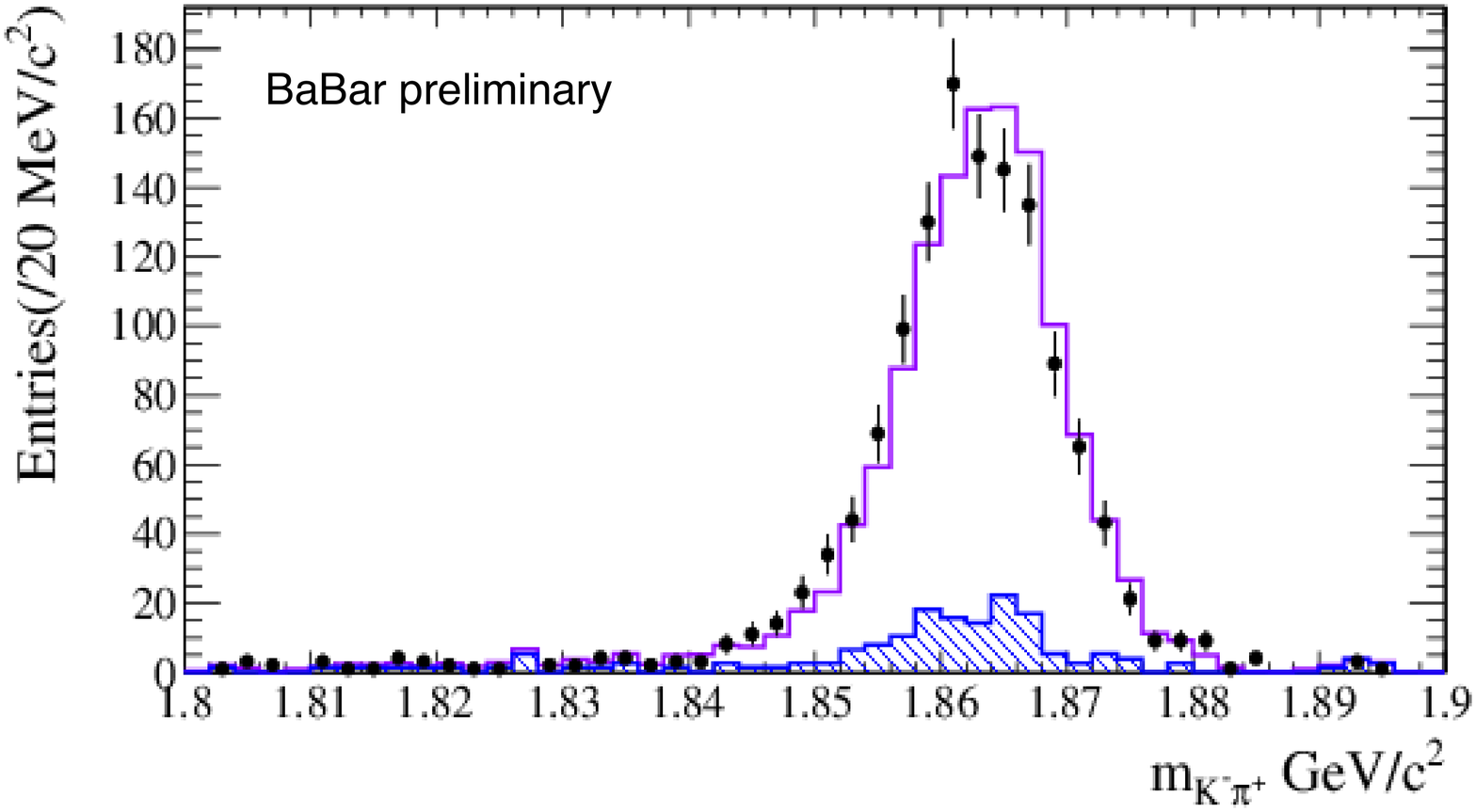}
\caption{MLP output distribution (left) for the three signal channels combined. The $B^+ \rightarrow K^+ \tau^+ \tau^-$ signal MC distribution is shown (dashed) with arbitrary normalization. The data (points) are overlaid on the  expected combinatorial (hatched) plus $m_{ES}$-peaking (solid line) background contributions. Invariant-mass distribution (right) of the $K^- \pi^+$ pair in $B^+ \rightarrow D^0 l^+ \nu_l$, $D^0 \rightarrow K^- \pi^+$  control samples after all signal selection criteria are applied, except for the final requirement on the MLP output.}
\label{fig1}
\end{figure*}

The predicted decay rate for $B^+ \rightarrow K^+ \tau^+ \tau^-$ in the SM is in the range $1-2 \times 10 ^{-3}$ \cite{bouchard, hewitt}. 
This decay is the third family equivalent of $B^+ \rightarrow K^+ l^+ l^-$, previously measured at \babar\ \cite{kll} and other experiments \cite{lhcb1},  
which shows some tension with the SM expectations \cite{klltheory}, and may provide additional sensitivity to new physics due to third-generation couplings and the large mass of the $\tau$ lepton. 
An important potential contribution to this decay is from neutral Higgs boson couplings, where the lepton-lepton-Higgs vertices are proportional to the squared mass of the leptons involved \cite{aliev}; thus, 
in the case of the $\tau$ lepton, such contributions can be significant and could alter the total decay rate. 
We use hadronic $B$ meson tagging techniques, where one of the two $B$ mesons, referred to as the $B_{tag}$, is reconstructed exclusively via its decay into one of several hadronic decay modes \cite{hadrec}. 
We consider only leptonic decays of the $\tau$, i.e. $\tau^+ \rightarrow e^+ \nu_e \nu_{\tau}$ and $\tau^+ \rightarrow \mu^+ \nu_{\mu} \nu_{\tau}$, which results in three different final states 
with an $ee$ , $\mu \mu$ or an $e \mu$ pair. 
Simulated Monte Carlo (MC) signal and background events, generated with EvtGen \cite{evtgen}, are used to develop signal selection criteria and to study potential backgrounds. 
We select $B_{tag}$ candidates using $\Delta E = \sqrt{s}/2 - E^*_{B_{tag}}$ and $m_{ES} = \sqrt{s - \vec p^{*2} _B}$, where $E^* _{B_{tag}}$ and $\vec p ^* _{B_{tag}}$ are the CM energy and three-momentum vector of the $B_{tag}$ respectively. 
We require a properly reconstructed $B_{tag}$ to have $m_{ES}$ consistent with the mass of a $B$ meson and $-0.12 < \Delta E < 0.12$ GeV. 
$B^+ \rightarrow K^+ \tau^+ \tau^-$ signal events are required to have a charged $B_{tag}$ candidate with $m_{ES} > 5.27$ GeV/c$^2$ and a non-zero missing energy, $E_{miss}$, given by the energy  component of $p^*_{miss}$. 
Continuum events are further suppressed using a multivariate likelihood selector, based on six event-shape variables 
which removes more than 75\% of the continuum events while retaining more than 80\% of (signal and background) $B \bar B$ MC events. 
Signal candidates are then required to possess exactly three charged tracks satisfying particle identification (PID) requirements consistent with one charged 
$K$ and an $e^+e^-$, $\mu^+ \mu^-$, or $e^+ \mu^-$ pair. 
Furthermore, events with $3.00 < m_{l^+ l^-} < 3.19$ GeV/c$^2$ are discarded to remove backgrounds from $J/\Psi$ resonance. 
The invariant mass of the combination of the $K$ with the oppositely charged lepton must also lie outside the region of the $D^0$ mass, i.e. $m_{K^-l^+} < 1.80$ GeV/c$^2$ or $m_{K^- l^+} > 1.90$ GeV/c$^2$, 
to remove events where a $\pi$ coming from the $D^0$ decay is misidentified as a muon. 
At this stage, remaining backgrounds are primarily $B \bar B$ events in which a properly reconstructed $B_{tag}$ accompanied by a $B_{sig} \rightarrow D^{(*)} l \nu_l$, with 
$D^{(*)} \rightarrow K l'\nu _{l'}$ which have the same detected final state particles as signal events. 
A multi-layer perceptron (MLP) neural network \cite{mlp}, with eight input variables and one hidden layer, is employed to suppress this background. 
The MLP is trained and tested using randomly split dedicated signal MC and $B^+ B^-$ background events, for each of the three channels. 
The results are shown in Fig.~\ref{fig1} (left) for the three modes combined. 
We require the output of the neural network is $>$ 0.70 for the $e^+ e^-$ and $\mu^+ \mu^-$ channels and $>$ 0.75 for the $e^+ \mu^-$ channel. 
This requirement is optimized to yield the most stringent upper limit in the absence of a signal.
A $B_{tag}$ yield correction is determined by calculating the ratio of data to $B^+ B^-$ MC events before the final MLP requirement. This correction factor is determined to be 
$0.913 \pm 0.020$ and is applied to the MC reconstruction efficiency for both signal and background events (Fig. \ref{fig1} (right)). 
The most important contributions to the systematic uncertainty include the uncertainty associated with the theoretical model which is evaluated by comparing signal MC sample based on the LCSR \cite{lcsr} theoretical model to that of \cite{mcp} and determining the difference in efficiency, which is found to be 3.0\%. 
Additional uncertainties on $\epsilon _{sig}$ and $N_{bkg}$ arise due to the modeling of PID selectors (4.8\% for $e^+ e^-$, 7.0\% for $\mu ^+ \mu ^-$, and 5.0\% for $e^+ \mu ^-$) and the $\pi ^0$ veto (3.0\%). 
The level of agreement between data and MC results in a systematic uncertainty of 2.6\%. 

The yields in the $e^+ e^-$ and $\mu^+ \mu^-$ channels are consistent with the expected background estimate. 
The signal yield in the $e^+ \mu^-$ channel is about twice the expected rate, which corresponds to an excess of $3.7 \sigma$ over the background expectation. 
Kinematic distributions in the $e^+ \mu^-$ do not give any clear hint of signal-like behavior or of systematic problems with background modeling. 
When combined with the $e^+ e^-$ and $\mu ^+ \mu^-$ modes, the overall significance of the $B^+ \rightarrow K^+ \tau^+ \tau^-$ signal is less than $2 \sigma$, and hence we 
do not interpret this as evidence of signal. 
Nevertheless, under the assumption that the excess observed is signal, the branching fraction for the combined three modes is $B(B^+ \rightarrow K^+ \tau ^+ \tau ^-) = (1.31^{+0.66} _{-0.61} ($stat.$) ^{+0.35} _{-0.25} ($sys.$)) \times 10^{-3}$. 
The upper limit at the 90\% confidence level is $B(B^+ \rightarrow K^+ \tau ^+ \tau ^-) <  2.25 \times 10^{-3}$.

\section{Angular analysis of $B \rightarrow K^* l^+ l^-$}
The amplitudes for the decays $B \rightarrow K^*(892) l^+ l^-$, where $K^* \rightarrow K \pi$ are expressed in terms of hadronic form factors and perturbatively-calculable effective Wilson coefficients, $C_7 ^{\rm eff}$ , $C_9 ^{\rm eff}$ and $C_{10} ^{\rm eff}$, which represent the electromagnetic penguin diagram, and the vector part and the axial-vector part of the linear combination of the $Z$ penguin and $W^+W^-$ box diagrams, respectively \cite{buchalla, altmann}. Non-SM physics may add new penguin and/or box diagrams, as well as possible contributions from new scalar, pseudoscalar, and/or tensor currents, which can contribute at the same order as the SM diagrams, modifying the effective Wilson coefficients from their SM expectations \cite{burdman, hewett}. 

\begin{figure*}[!htb]
\centering
\includegraphics[width=0.85\textwidth]{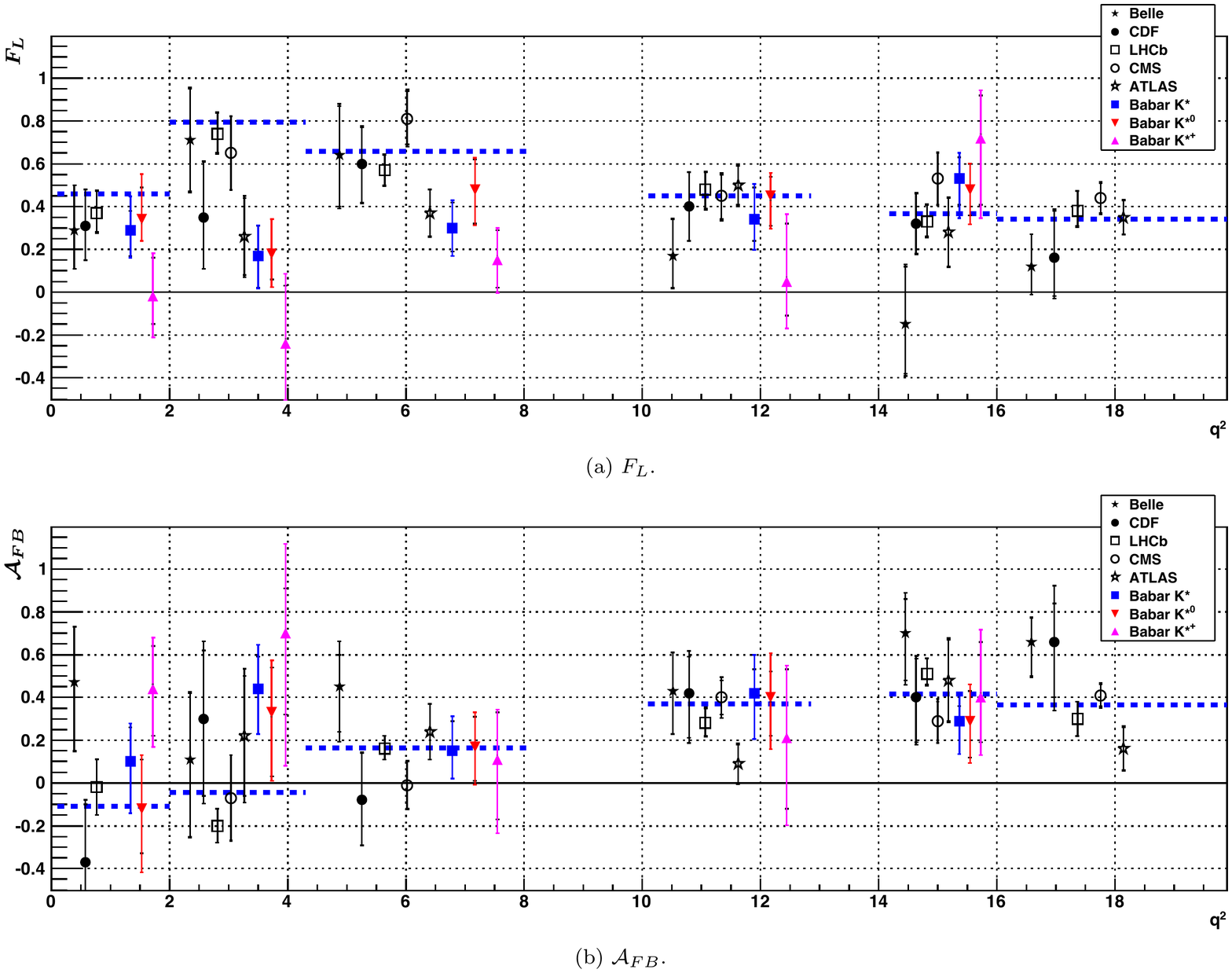}
\caption{$F_L$ (top) and $\mathcal A_{FB}$ (bottom) results in disjoint $q^2$ bins, along with those of other experiments and the SM expectations (blue dashed lines, which also define the extent of each individual 
$q^2$ bin): (black filled star) Belle \cite{belle}, (black filled circle) CDF \cite{cdf}, (black open square) \cite{lhcb}, (black open circle) CMS \cite{cms}, (black open star) ATLAS \cite{atlas}, (blue filled square) 
\babar\ $B \rightarrow K^* l^+ l^-$, (red filled down-pointing triangle) $B^0 \rightarrow K^{*0} l^+ l^-$ , (magenta filled up-pointing triangle) $B^+ \rightarrow K^{*+} l^+ l^-$. The \babar\ $q _5 ^2$ results are 
drawn in the $14 < q^2 < 16$ GeV$^2$/c$^4$ region, however, they are valid for the entire $q^2 > 14$ GeV$^2$/c$^4$ region.
}
\label{fig7}
\end{figure*}

The angular distributions in $B \rightarrow K^* l^+ l^-$ decays, as function of squared di-lepton mass $q^2 = m^2 _{l^+ l^-}$, are sensitive to many new physics models, with several measurements presented over the past few years \cite{belle}-\cite{atlas}. 
For a given $q^2$ value, the kinematic distribution of the decay products 
can be expressed as a triply differential cross-section in three angles: $\theta_K$ , the angle between the $K$ and the $B$ directions in the $K^*$ rest frame; 
$\theta_l$, the angle between the $l^+$ and the $B$ direction in the $l^+l^-$ rest frame; and $\phi$, the angle between the $l^+l^-$ and $K \pi$ decay planes in the $B$ rest frame. 
From the distribution of the angle $\theta _K$ obtained after integrating over $\phi$ and $\theta _l$ , we determine the $K^*$ longitudinal polarization fraction $F_L$ using a fit to $\cos \theta _K$ of the form 
\begin{equation}
\frac{1}{\Gamma (q^2)}\frac{d\Gamma} {d(\cos \theta _K)} = \frac{3}{2} F _L(q^2) \cos \theta _K + \frac{3}{4} (1 - F _L(q ^2))(1 - \cos^2 \theta _K)
\end{equation}
while, integrating over $\phi$ and $\theta _K$ we extract the forward-backward asymmetry $\mathcal A _{FB}$ from a fit to $\cos \theta _l$ 
\begin{equation}
\frac{1}{\Gamma (q^2)}\frac{d\Gamma} {d(\cos \theta _l)}= \frac{3}{4} F_L(q^2)(1 - \cos^2 \theta_l) + \frac{3}{8}(1 - F_L(q^2))(1 + \cos^2 \theta _l) + \mathcal A _{FB}(q^2) \cos \theta _l . 
\end{equation}
We determine $F_L$ and $\mathcal A _{FB}$ in the five disjoint bins of $q^2$ (Fig.~\ref{fig7}). We also present results in a $q^2$ range $1.0 < q_0 ^2 < 6.0$ GeV$^2$/c$^4$, the perturbative window away 
from the $q^2 \rightarrow 0$ photon pole and the $c \bar c$ resonances at higher $q^2$, where theory uncertainties are considered to be under good control.
We reconstruct signal events in 5 different final states: $B^+ \rightarrow K^{*+}(\rightarrow K_S ^0 \pi^+) \mu^+ \mu^-$, $B^0 \rightarrow K^{*0}(\rightarrow K^+ \pi^-) \mu^+ \mu^-$, $B^+ \rightarrow K^{*+}(\rightarrow K^+ \pi ^0)e^+ e^-$, $B^+ \rightarrow K^{* +} (\rightarrow K_S ^0 \pi ^+) e^+ e^-$  and $B^0 \rightarrow K^{*0}(\rightarrow K^+ \pi^-)e^+ e^-$; we do not include other modes as their signal/background ratio is seen to be very poor. 
We require $K^*$ candidates to have an invariant mass $0.72 < m_{K\pi} < 1.10$ GeV/c$^2$. 
We reconstruct $K_S ^0$ candidates in the $\pi ^+ \pi ^-$ final state, requiring an invariant mass consistent with the nominal $K ^0$ mass, and a flight distance from the $e^+ e^-$ interaction 
point that is more than three times the flight distance uncertainty. Neutral pion candidates are formed from two photons with $E_{\gamma} > 50$ MeV, and an invariant mass 
between 115 and 155 MeV/c$^2$. In each final state, we use the kinematic variables $m_{ES}$ and $\Delta E$ as defined in the previous section. We reject events with $m_{ES} < 5.2$ GeV/c$^2$.

Random combinations of leptons from semileptonic $B$ and $D$ decays are the predominant source of backgrounds; these combinatorial backgrounds occur in both 
$B \bar B$ events 
and $e^+e^- \rightarrow q \bar q$ continuum events 
(where $q = u,d,s,c$), and are suppressed using eight bagged decision trees (BDTs) \cite{bdt} depending on the background class, final state ($e e$ or $\mu \mu$), and $q^2$ region. 

We extract the angular observables $F_L$ and $\mathcal A_{F B}$ from the data using a series of likelihood (LH) fits which proceed in several steps:
\begin{itemize}
\item{In each $q^2$ bin, for each of the five signal modes separately and using the full $m_{ES} > 5.2$ GeV/c$^2$ dataset, an initial unbinned maximum LH fit of $m_{ES}$, $m_{K \pi}$ and a likelihood ratio 
that discriminates against random combinatorial $B \bar B$ backgrounds is performed. After this first fit, all normalizations and the probability density function (pdf) shapes are fixed.} 
\item{Second, in each $q^2$ bin and for each of the five signal modes, $m_{ES}$, $m_{K \pi}$ and LR pdfs and normalizations are defined for $m_{ES} > 5.27$ GeV/c$^2$ events (the ``$m_{ES}$ angular fit region'') 
using the results of the prior three-dimensional fits. Only $m_{ES}$ angular fit region events and pdfs are subsequently used in the fits for $F_L$ and $\mathcal A_{FB}$.}
\item{Next, $\cos \theta _K$ is added as fourth dimension to the likelihood function, and four-dimensional likelihoods with $F_L$ as the only free parameter are defined for $m_{ES}$ angular fit region events. 
Each $q^2$ bin and each of the five signal modes has its own separate LH function. Thus, it becomes possible to extract $F_L$ and $\mathcal A_{FB}$ for arbitrary combinations of the five final states. In particular, we 
quote results using three different sets of our five signal modes: the charged mode $B^+ \rightarrow K^{*+} l^+ l^-$, the neutral mode $B^0 \rightarrow K^{*0} l^+ l^-$, and the inclusive mode.}
\item{
In the final step, we use the fitted value of $F_L$ from the previous fit step as input to a similar 4-d fit for $\mathcal A_{FB}$, in which $\cos \theta _l$ replaces $\cos \theta _K$ as the fourth dimension in the LH function.}
\end{itemize}

Fig.~\ref{fig7} graphically shows our $F_L$ and $\mathcal A_{F B}$ results \cite{paper2} in disjoint $q^2$ bins alongside other published results and the SM theory expectations, 
the latter of which typically have 5-10\% theory uncertainties (absolute) in the regions below and above the charmonium resonances. 


\section{Study of $B \rightarrow K \pi^+ \pi^- \gamma$ decays}

\begin{figure*}[!htb]
\centering
\includegraphics[width=0.47\textwidth]{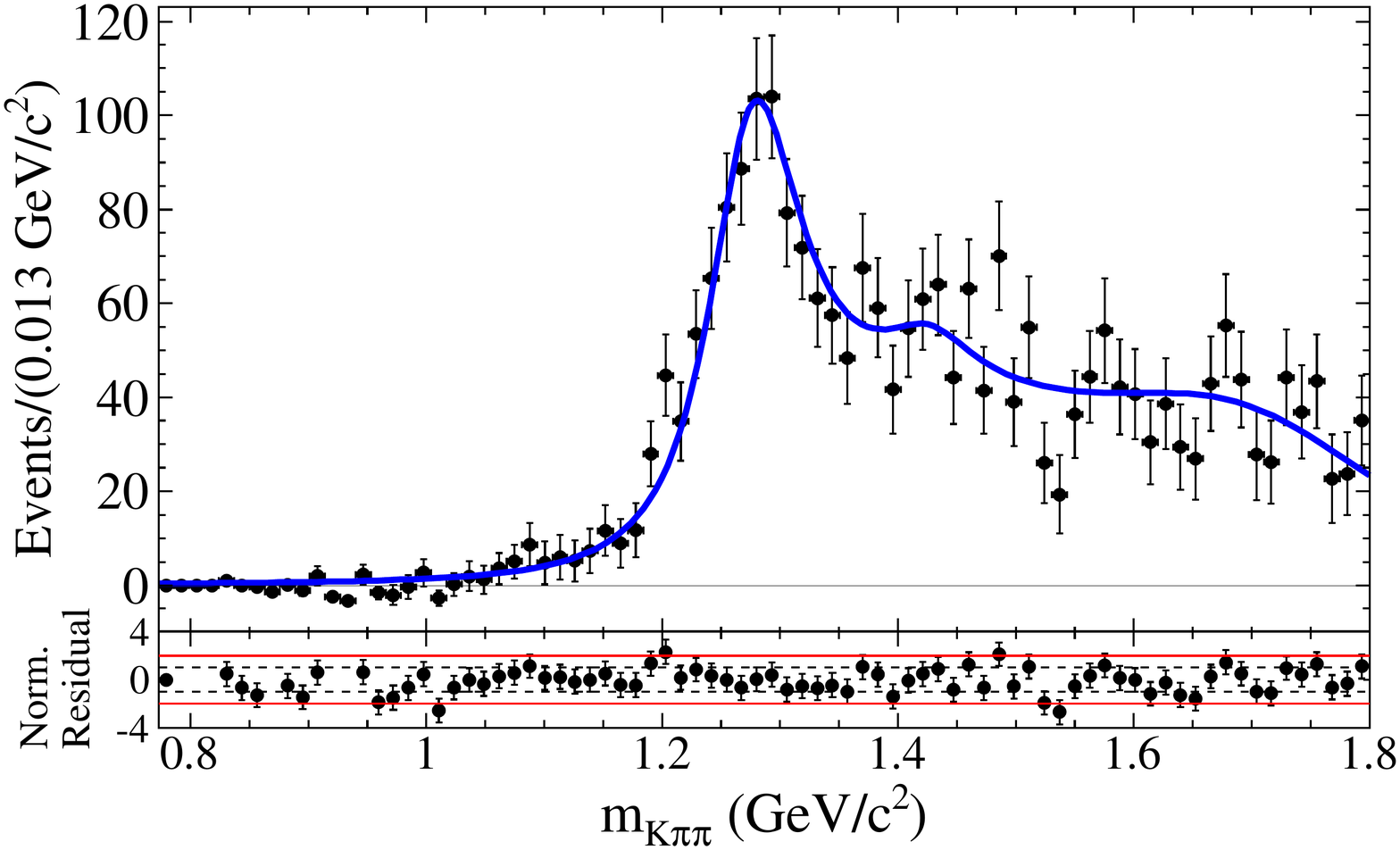}
\quad
\includegraphics[width=0.48\textwidth]{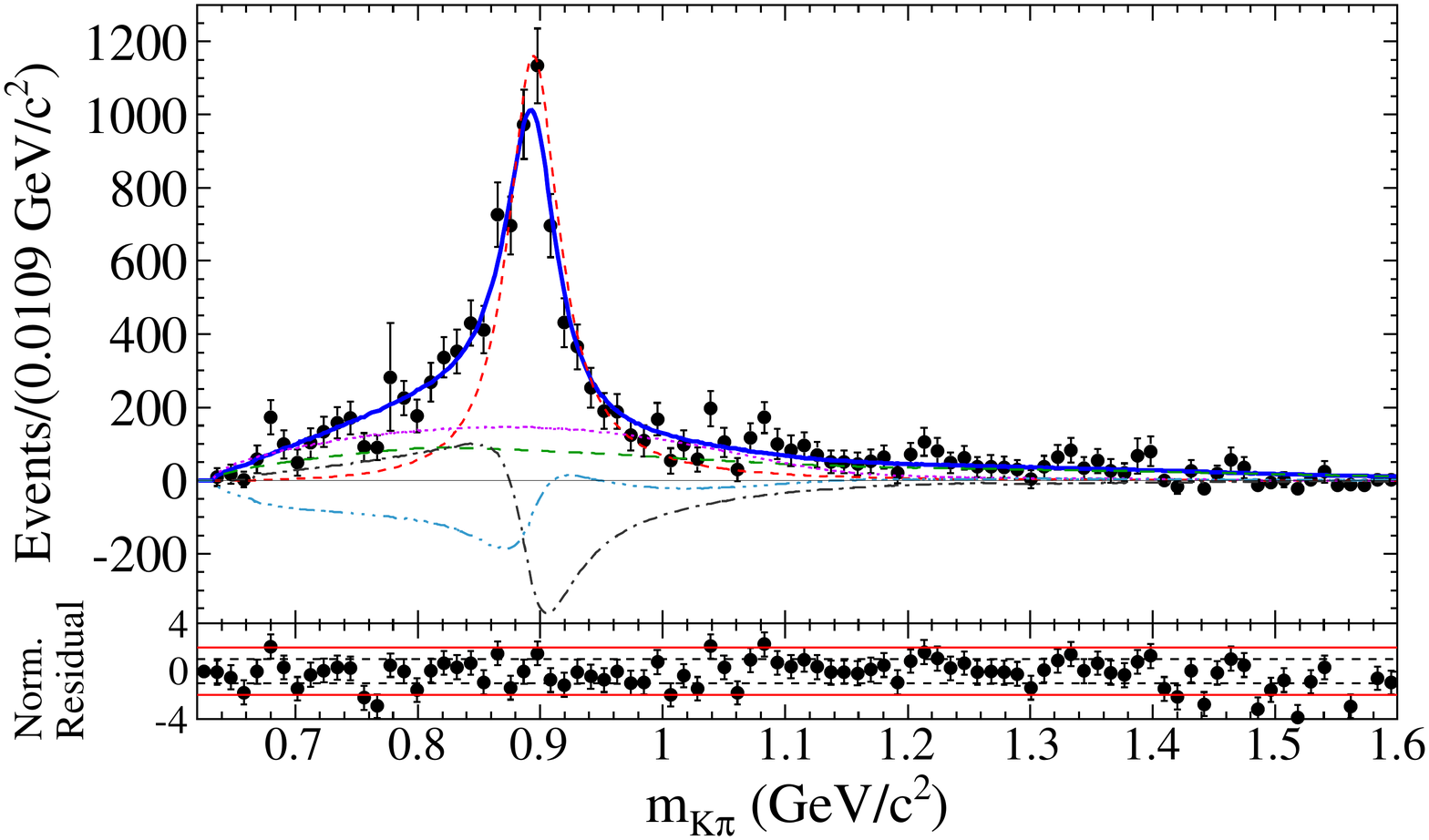}
\caption{The $m_{K \pi \pi}$ (left) and $m_{K \pi}$ (right) spectra for correctly-reconstructed $B^+ \rightarrow K^+ \pi^- \pi^+ \gamma$ signal 
events extracted from maximum likelihood fit to $m_{ES}$, $\Delta E$ and $\mathcal F$ using $_s \mathcal P lot$ technique. Points with error bars give the sum of sWeights. 
The blue solid curve shows the fit to the $m_{K \pi \pi}$ spectrum and the total PDF fit projection to $m_{K \pi}$, respectively. 
The small-dashed red, medium-dashed green and dotted magenta curves correspond to the $K^*(892) ^0$, $\rho(770) ^0$ and $(K \pi)^{*} _0$ contributions, respectively. 
The dashed-dotted gray curve corresponds to the interference between the two P-wave components, i.e. the $K^*(892) ^0$ and the $\rho (770) ^0$, 
and the dashed-triple-dotted light blue curve corresponds to the interference between the $(K \pi)^{*} _0$ and the $\rho (770) ^0$. 
Below each bin are shown the residuals, normalized in units of standard deviations, where the parallel dotted and full lines mark the one and two standard deviation levels, respectively.}
\label{fig2}
\end{figure*}

The V-A structure of the SM weak interaction implies that the circular polarization of the photon emitted in $b \rightarrow s \gamma$ transitions is predominantly left-handed, 
with contamination by right-handed photons suppressed by a factor $m_s/m_b$. Thus, $B^0$ mesons decay mostly to right-handed photons while decays of $\bar B^0$ mesons produce mainly left-handed 
photons. Therefore, the mixing-induced CP asymmetry in $B \rightarrow f_{CP} \gamma$ decays, where $f_{CP}$ is a CP eigenstate, is expected to be small. 
This prediction may be altered by new-physics (NP) processes in which opposite helicity photons are involved. Especially, in some NP models \cite{fuji}, the right-handed component may be comparable in 
magnitude to the left-handed component, without affecting the SM prediction for the inclusive radiative decay rate. 
Our goal consists of measuring the mixing-induced CP asymmetry parameter, $S_{K^0 _S \rho^0 \gamma}$, in the radiative B decay to the CP eigenstate $K_S ^0 \rho ^0 \gamma$, which is sensitive to 
right-handed photons. Because of the irreducible background from $B^0 \rightarrow K_S ^0 \pi^+ \pi^- \gamma$, which is not a CP eigenstate, we have to measure first the time-dependent CP asymmetry parameters $S_{K^0 \pi \pi \gamma}$ and $C_{K^0 \pi \pi \gamma}$ which are related to $S_{K^0 _S\rho^0 \gamma}$ by $S_{K^0 _S \rho^0 \gamma} = S_{K^0 \pi \pi \gamma} / D _{K_S ^0 \rho ^0 \gamma}$
where the dilution factor $D _{K_S ^0 \rho ^0 \gamma}$ depends on the amplitudes of the two-body decays $K_S ^0 \rho(770)^0 $, $K^* (892)^+ \pi^-$ and $(K\pi)^+ _0 \pi ^-$ and can be calculated as shown in \cite{dilution}. 
Because of the much higher signal yield for $B^+ \rightarrow K^+ \pi^+ \pi^- \gamma$, compared to the neutral mode, in our work the dilution factor $D_{K_S ^0 \rho ^0 \gamma}$ is determined from a study of the charged mode $B^+ \rightarrow K^+ \pi^+ \pi^- \gamma$, which is related to the neutral mode by isospin symmetry.
Using an extended unbinned maximum likelihood fit to $m_{ES}$, $\Delta E$ and the Fisher discriminant $\mathcal F$ \cite{fisher}, 
we extract the signal yield for $B^+ \rightarrow K^+ \pi^+ \pi^- \gamma$ for $m_{K \pi \pi} < 1.8$ GeV/c$^2$. Using the sPlot technique \cite{splot}, 
we extract the $m_{K \pi \pi}$, $m_{K \pi}$ and $m_{\pi \pi}$ invariant-mass spectra. We model the $m_{K \pi \pi}$ invariant-mass spectrum 
as coherent sum of five resonances (Fig.~\ref{fig2}), each parameterized by a relativistic Breit-Wigner line shape and from a maximum likelihood fit to the $m_{K \pi \pi}$ spectrum we derive the branching fractions of 
the individual kaonic resonances shown in Table~\ref{tabella1}. 
We measure the $B^+ \rightarrow K^+ \pi^+ \pi^- \gamma$ branching fraction to be $B(B^+ \rightarrow K^+ \pi^+ \pi^- \gamma) = (27.2\pm1.0\pm1.2) \times 10^{-6}$ \cite{paper3}. 

\begin{table}[!htb]
\begin{center}
\caption{Branching fractions of the different $K^+ \pi ^- \pi ^+$ resonances extracted from the fit to the $m_{K \pi \pi}$ spectrum. To correct for the secondary branching fractions, we use the values in PDG \cite{pdg04}. The first uncertainty is statistical, the second is systematic, and the third, when present, is due to the uncertainties on the secondary branching fractions. ``n/a'' indicates that the corresponding branching fraction was not previously reported.}
\begin{tabular}{cccc}
Mode & \begin{tabular}[c]{@{}c@{}}$B(B^+ \rightarrow$ Mode) $\times$ \\$B(K_{res} \rightarrow K^+ \pi^+ \pi^-) \times 10^{-6}$\end{tabular}
& $B(B^+ \rightarrow$ Mode) $\times 10^{-6}$ & \begin{tabular}[c]{@{}c@{}} Previous world \\ average $(\times 10^{-6})$ \end{tabular} \\
\hline
\hline
$B^+ \rightarrow K^+ \pi ^+ \pi ^- \gamma$ & ... & $24.5 \pm 0.9 \pm 1.2$ & $27.6 \pm 2.2$ \\
\hline
$K_1 (1270) ^+ \gamma$ & $14.5^{+2.1 +1.2} _{-1.4 -1.2}$ & $44.1^{+6.3 +3.6} _{-4.4 -3.6} \pm 4.6$ & $43 \pm 13$ \\
$K_1 (1400) ^+ \gamma$ & $4.1^{+1.9 +1.2} _{-1.2 -1.0}$ & $9.7^{+4.6 +2.8} _{-2.9 -2.3} \pm 0.6$ & $<$ 15 at 90\% CL \\
$K^* (1410) ^+ \gamma$ & $11.0^{+2.2 +2.1} _{-2.0 -1.1}$ & $27.1^{+5.4 +5.2} _{-4.8 -2.6} \pm 2.7$ & n/a \\
$K_2 ^* (1430) ^+ \gamma$ & $1.2^{+1.0 +1.2} _{-0.7 -1.5}$ & $8.7^{+7.0 +8.7} _{-5.3 -10.4} \pm 0.4$ & $14 \pm 4$ \\
$K^* (1680) ^+ \gamma$ & $15.9^{+2.2 +3.2} _{-1.9 -2}$ & $66.7^{+9.3 +13.3} _{-7.8 -10.0} \pm 5.4$ & $<$ 1900 at 90\% CL \\
\hline
\end{tabular}
\end{center}
\label{tabella1}
\end{table}

We then perform a further maximum likelihood fit to the $m_{K \pi}$ spectrum in which we include $\rho (770) ^0$, $K^* (892) ^0$ and a $(K^+ \pi ^-)_0$ non-resonant S-wave contribution. 
We model the $K^*(892) ^0$ with a relativistic Breit-Wigner line shape, the $\rho(770) ^0$ with a Gounaris-Sakurai line shape and the $(K^+ \pi^-)_0$ with the LASS parameterization \cite{lass}. 
Table~\ref{tabella2} lists the branching fraction of the different resonances decaying to $K^+ \pi ^-$ and $\pi ^+ \pi^-$. 
This is the first observation of the decay $B^+ \rightarrow K^+ \rho ^0 \gamma$ and the $B^+ \rightarrow (K \pi)^{*0} _0 \pi ^+ \gamma$ S-wave contribution.
From the measured two-body amplitudes we obtain a dilution factor of $D _{K_S ^0 \rho \gamma} = -0.78 ^{+0.19} _{-0.17}$ with mass constraints 
$m_{K \pi \pi} < 1.8$ GeV/c$^2$, $0.6 < m_{\pi \pi} < 0.9$ GeV/c$^2$, $m_{K \pi} < 0.845$ GeV/c$^2$ and $m_{K \pi} > 0.945$ GeV/c$^2$.

\begin{figure*}[!htb]
\centering
\includegraphics[width=.48\textwidth]{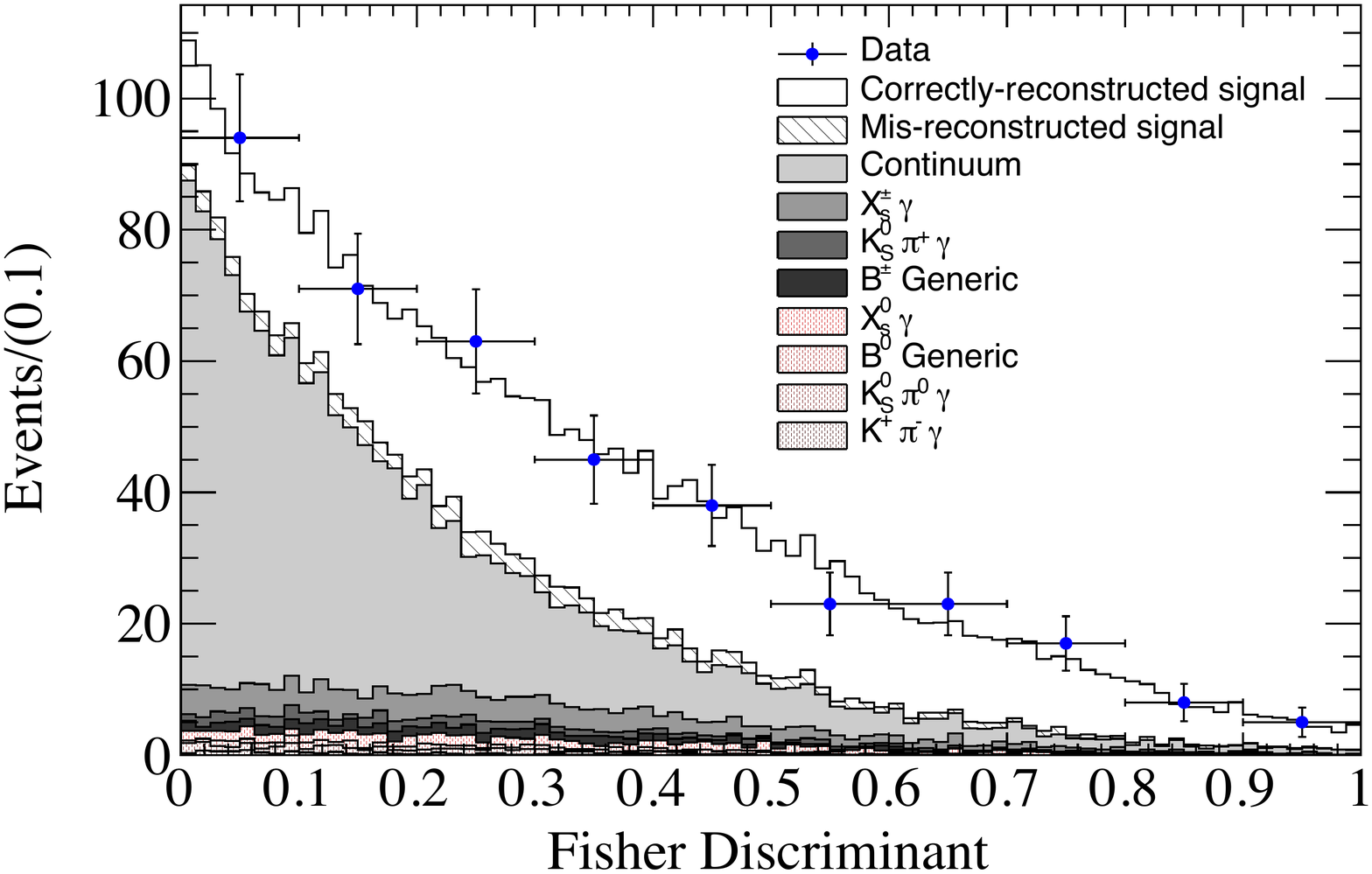}
\quad
\includegraphics[width=.485\textwidth]{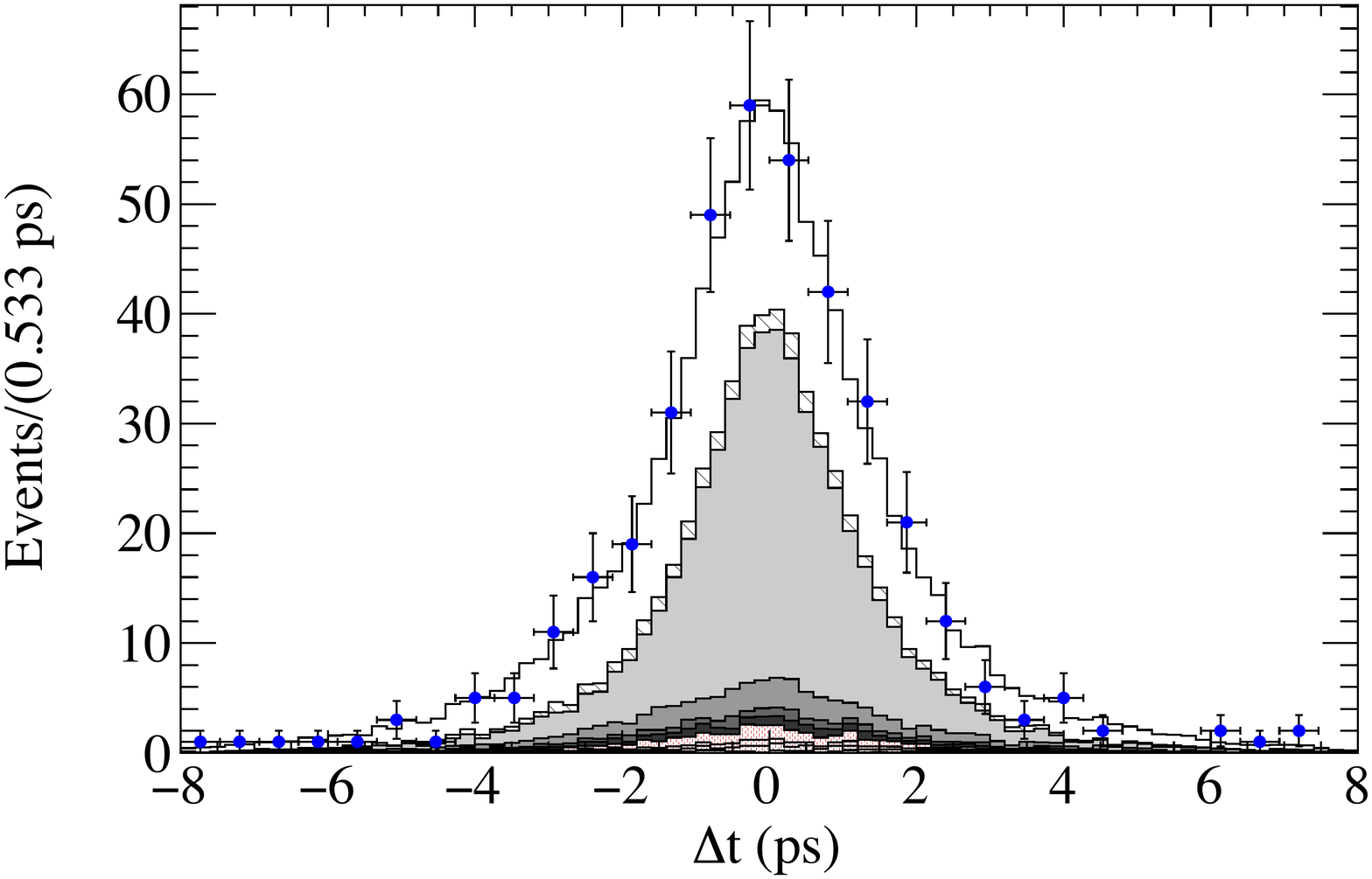}
\caption{Distributions of Fisher discriminant (left) and $\Delta t$ (right), showing fit results to the $B^0 \rightarrow K_S ^0 \pi^+ \pi^- \gamma$ data sample with the additional requirements: 
$-0.15 \le \Delta E \le 0.10$ GeV $(m_{ES})$, $m_{ES} > 5.27$ GeV/c$^2$ and $-0.15 \le \Delta E \le 0.10$ GeV ($\Delta t$). 
Points with error bars show the data. The projection of the fit result is represented by stacked histograms, where the shaded areas represent the background contributions, as described in the legend. 
}
\label{fig3}
\end{figure*}

\begin{table}[!htb]
\begin{center}
\caption{Branching fractions of the resonances decaying to $K \pi$ and $\pi \pi$ extracted from the fit to the $m_{K\pi }$ spectrum. R denotes an intermediate resonant state and $h$ stands for a final state hadron: a charged pion or kaon. To correct for the secondary branching fractions, we use values from \cite{pdg04} and $B(K^*(892)^0 \rightarrow K^+ \pi^-) = 2$. The first uncertainty is statistical, the second is 
systematic, and the third (when applicable) is due to the uncertainties on the secondary branching fractions. 
The last two rows of the table are obtained by separating the contributions from the resonant and the nonresonant part of the LASS parametrization. ``n/a'' indicates that the corresponding branching fraction was not previously reported.}
\begin{tabular}{cccc}
 & & & \\
Mode & \begin{tabular}[c]{@{}c@{}}$B(B^+ \rightarrow$ Mode) $\times$ \\$B(R \rightarrow h \pi) \times 10^{-6}$\end{tabular}
& $B(B^+ \rightarrow$ Mode) $\times 10^{-6}$ & \begin{tabular}[c]{@{}c@{}} Previous world \\ average $(\times 10^{-6})$ \end{tabular} \\
\hline
\hline
$K^* (892)^0 \pi^+ \gamma$ & $15.6 \pm 0.6 \pm 0.5$ & $23.4 \pm 0.9 ^{+0.8} _{-0.7}$ & $20 ^{+7} _{-6}$ \\
$K^+ \rho (770)^0 \gamma$ & $8.1 \pm 0.4^{+0.8} _{-0.7}$ & $8.2 \pm 0.4 \pm 0.8 \pm 0.02$ 
& $<$ 20 at 90\% CL\\
$(K \pi ) ^{*0} _0 \pi^+ \gamma$ & $10.3 ^{+0.7+1.5} _{-0.8 -2.0}$ & ... & n/a \\
\hline
$(K \pi )^0 _0 \pi^+ \gamma$ (NR) & ... & $9.9 \pm 0.7 ^{+1.5} _{-1.9}$ & $<$ 9.2 at 90\% CL \\
$K_0 ^*(1430)^0 \pi^+ \gamma$ & $0.82 \pm 0.06 ^{+0.12} _{-0.16}$ & $1.32 ^{+0.09+0.20} _{-0.10-0.26} \pm 0.14$ & n/a \\
\hline
\end{tabular}
\end{center}
\label{tabella2}
\end{table}

Finally for $B^0 \rightarrow K^0 _S \pi^+ \pi^- \gamma$ we use a selection similar to $B^+ \rightarrow K^+ \pi^+ \pi^- \gamma$ and by means of a maximum likelihood fit we extract signal yield $B(B^0 
\rightarrow K^0 \pi^+ \pi^- \gamma) = (24.0\pm2.4 ^{+1.7} _{-1.8}) \times 10^ {-6}$ and CP asymmetry parameters $S_{K_S ^0 \pi^+ \pi^- \gamma} = 0.14 \pm 0.25 \pm 0.03$ 
and $C_{K_S ^0 \pi^+ \pi^- \gamma}  = -0.39 \pm 0.20 ^{+0.03} _{-0.02}$ from which we finally get $S_{K_S ^0 \rho ^0 \gamma} = -0.18 \pm 0.32 ^{+0.06} _{-0.05}$ in agreement with the SM. 

\section{Conclusion}
In conclusion, we performed the first search for the decay $B^+ \rightarrow K^+ \tau^+ \tau^-$; 
no significant signal is observed and the upper limit on the final branching fraction is determined to be $2.25 \times 10^{-3}$ at the 90\% confidence level. 
We have measured the fraction $F_L$ of longitudinally polarized $K^*$ decays and the lepton forward-backward asymmetry $\mathcal A_{FB}$ in bins of dilepton mass-squared in 
$B^+ \rightarrow K^{*+} l^+ l^-$, $B^0 \rightarrow K^{*0} l^+ l^-$ and $B \rightarrow K^* l^+ l^-$. Results for the charged mode are presented for the first time here. 
Our $B^0 \rightarrow K^{*0} l^+ l^-$ results are in reasonable agreement with both SM theory expectations and other experimental results. 
Similarly, although with relatively larger uncertainties, we observe broad agreement of the $B^+ \rightarrow K^{*+} l^+ l^-$ results with those for $B^0 \rightarrow K^{*0} l^+ l^-$. 
However, in the low dilepton mass-squared region, we observe relatively very small values for $F_L$ in $B^+ \rightarrow K^{*+} l^+ l^-$, exhibiting tension with 
both the $B^0 \rightarrow K^{*0} l^+ l^-$ results as well as the SM expectations. 
These tensions in $F_L$ are difficult to interpret because of uncertainties due to form-factor contributions in the calculation of this observable in both the SM and NP scenarios. 
We measured the branching fractions of the decays $B^+ \rightarrow K^+ \pi^+ \pi^- \gamma$ and $B^0 \rightarrow K_S ^0 \pi^+ \pi^- \gamma$. For $B^+ \rightarrow K^+ \pi^+ \pi^- \gamma$ 
we observed five different resonances decaying to $K \pi \pi$ state and we measured their branching fractions. 
We found first evidence for $B \rightarrow K_1(1400) ^+ \gamma$, $B \rightarrow K^*(1410) ^+ \gamma$ and $B \rightarrow K^*(1400) ^+ \gamma$ decays. 
We have calculated the dilution factor $D_{K_S ^0 \rho ^0 \gamma}$ from the measurement of $B \rightarrow K^*(892) ^0 \pi^+ \gamma$, 
$B^+ \rightarrow K ^+ \rho(770) \gamma$ and $B \rightarrow (K ^+ \pi^-) ^0 \pi ^+ \gamma$ decays. 
We have measured the time-dependent CP asymmetry parameters $S_ {K _S ^0 \pi ^+ \pi ^- \gamma}$ and 
$C_{K _S ^0 \pi^+ \pi- \gamma}$ and hence derived $S_{K _S ^0 \rho ^0 \gamma}$ for the $K _S ^0 \rho(770) ^0 \gamma$ CP eigenstate.

\end{document}